# Nutritionally recommended food for semi- to strict vegetarian diets based on large-scale nutrient composition data

Seunghyeon Kim[1,2,3], Michael F. Fenech[4] & Pan-Jun Kim[1,2,5,6,*]

[1]Department of Physics, Pohang University of Science and Technology, Pohang, Gyeongbuk 37673, Republic of Korea
[2]Asia Pacific Center for Theoretical Physics, Pohang, Gyeongbuk 37673, Republic of Korea
[3]The Abdus Salam International Centre for Theoretical Physics, 34151 Trieste, Italy
[4]Genome Health Foundation, North Brighton, South Australia 5048, Australia
[5]Department of Physics, Korea Advanced Institute of Science and Technology, Daejeon 34141, Republic of Korea
[6]Department of Biology, Hong Kong Baptist University, Kowloon, Hong Kong
*Correspondence and requests for materials should be addressed to P.-J.K. (email: panjunkim@hkbu.edu.hk)

**Diet design for vegetarian health is challenging due to the limited food repertoire of vegetarians. This challenge can be partially overcome by quantitative, data-driven approaches that utilise massive nutritional information collected for many different foods. Based on large-scale data of foods' nutrient compositions, the recent concept of nutritional fitness helps quantify a nutrient balance within each food with regard to satisfying daily nutritional requirements. Nutritional fitness offers prioritisation of recommended foods using the foods' occurrence in nutritionally adequate food combinations. Here, we systematically identify nutritionally recommendable foods for semi- to strict vegetarian diets through the computation of nutritional fitness. Along with commonly recommendable foods across different diets, our analysis reveals favourable foods specific to each diet, such as immature lima beans for a vegan diet as an amino acid and choline source, and mushrooms for ovo-lacto vegetarian and vegan diets as a vitamin D source. Furthermore, we find that selenium and other essential micronutrients can be subject to deficiency in plant-based diets, and suggest nutritionally-desirable dietary patterns. We extend our analysis to two hypothetical scenarios of highly personalised, plant-based methionine-restricted diets. Our nutrient-profiling approach may provide a useful guide for designing different types of personalised vegetarian diets.**

**Introduction**

Recent lifestyle changes represented by the high-fat/high-sugar 'Western' diet have raised public concern, as they contribute to the growing epidemics of metabolic and cardiovascular diseases in developed and developing countries worldwide[1–3]. On the other hand, a high consumption of plant-derived foods such as fruits, vegetables, nuts, and seeds is known to be associated with a lower risk of chronic diseases and cancer[4–6]. In addition, cultural background, as well as environmental or ethical motivations against meat production, becomes a favourable factor for adoption of plant-based or vegetarian diets[7, 8]. However, the strict practice of vegetarian diets is not devoid of drawbacks, as exemplified by a vegan diet that excludes all animal-derived foods (such as eggs and dairy products), which results in a high risk of vitamin $B_{12}$ deficiency because this vitamin is mainly available in animal products[9]. Other essential micronutrients that may be deficient in unsupplemented vegans include vitamin D, ω-3 fatty acids, calcium, zinc, and iron[10]. Therefore, a clear need exists for the identification of foods recommended for vegetarian health[11, 12], which is a challenging task due to the narrower food repertoire of vegetarians.

Large-scale, data-driven analysis methods are now widely used for fundamental quantitative inquiries into various biological, technological, and social systems[13–17]. These techniques have even been applied to foods[18, 19]. For example, a seminal analysis of a network connecting various food ingredients to flavour compounds revealed unforeseen regional variations in culinary cultures[19]. Recently, using nutritional information from over 1,000 raw foods, we have systematically evaluated the nutrient composition of each food with regard to satisfying the daily nutritional requirements, and thereby quantified the food's nutrient balance, or *nutritional fitness*[20]. Nutritional fitness offers an objective way to prioritise recommended foods for each physical and dietary condition (for example, physical conditions include ages, genders, heights, weights, and physical activity levels; see Supplementary Methods). Although scoring foods based on nutrient contents has also been attempted in other previous studies[21, 22], their methods generally involve rather arbitrarily structured mathematical formulas and explicit weighting factors of individual nutrients, which may lead to biased results. In contrast, the computation of nutritional fitness takes a conceptually different and clearly defined approach, based on the food's frequency of occurrence in nutritionally adequate food combinations that simultaneously constrain all nutrient levels within the ranges recommended for daily intake (see below). It should be noted that the broad notion of optimisation-based diet design is dating back to the historic use of the simplex method[23].

Here, we explore a spectrum of semi- to strict vegetarian diets, with regard to food recommendations for those diets based on nutritional fitness. Our study identifies both nutritionally recommendable foods and key relevant nutrients in each vegetarian diet. To demonstrate the applicability of our method to personalised nutrition[24, 25], we further investigate two hypothetical scenarios of semi-vegetarian, methionine-restricted diets (a methionine-restricted diet is a predominantly plant-based diet, with growing evidence of its metabolically unique health benefits, such as anti-cancer and healthy ageing effects[26]). The results from this study not only help improve our basic understanding of the nutritional structure of diets with specific food choices, but also have a wide range of implications in personalised nutrition, nutritional policies, and the food industry.

**Methods**

**Food and nutritional data and various diets**

We adopted information on the nutrient contents and energy (calorie) densities of foods (quantities per 100 g for each food) in our previous study[20], which was collected from the United States Department of Agriculture (USDA) National Nutrient Database for Standard Reference, Release 24[27]. In this previous study, we considered only raw foods and other foods whose nutrient contents have been minimally modified. Specifically, we selected foods in their natural forms without any explicitly added or fortified ingredients (e.g., added salt, sugar, and vitamins). In addition, we chose similar foods that were altered to ground, frozen, dried, low-fat, non-fat, or ultraviolet-treated products. In total, 1,068 foods were selected, and we refer to them as raw foods. A systematic unification of the raw foods redundant in their nutrient contents[20] yielded a total of 653 foods (Supplementary Methods).

The categories of foods were identified through the average-linkage hierarchical clustering of foods that are similar in their nutrient compositions, as described in our previous study[20]. As a result, most foods were divided into four major categories, roughly based on the foods' relative macronutrient levels[20]: protein-rich, fat-rich, carbohydrate-rich, and low-macronutrient categories (Supplementary Data S1). Accordingly, plant-derived foods mainly belong to the carbohydrate-rich, fat-rich, and low-macronutrient categories, and animal-derived foods mainly belong to the protein-rich and fat-rich categories.

In this study, we initially consider the following four diet styles of a physically active, 20-year-old male with standard height and weight (Supplementary Data S1): (*i*) control diet, which is a typical omnivorous diet that includes all 653 animal- and plant-derived foods, (*ii*) ovo-lacto vegetarian diet, which includes only 449 plant-derived foods, eggs, and dairy products, (*iii*) vegan diet, which includes only 434 plant-derived foods without any animal products, and (*iv*) methionine-restricted diet, which includes all 653 animal- and plant-derived foods but allows only a minimally required intake of methionine. Therefore, in practice, diet (*iv*) is predominantly vegetarian, as plant-derived foods tend to have a lower methionine content than animal-derived foods[26]. In addition, we later consider the following two hypothetical scenarios of highly personalised, methionine-restricted diets with 532 foods (Supplementary Data S2): (*v*) 61-year-old male with low physical activity and (*vi*) physically active 58-year-old female. Specifically, diets (*v*) and (*vi*) include the plant-derived foods in diet (*iii*), along with eggs [≤2 eggs/week for diet (*v*) and ≤1 egg/week for diet (*vi*)], whole milk [≤150 ml/week for diet (*v*) and ≤50 ml/week for diet (*vi*)], certain types of fish [≤150 g/week for diet (*v*) and ≤100 g/week for diet (*vi*)], and cheese [≤40 g/week for diet (*v*) and ≤20 g/week for diet (*vi*)], although cheese products do not belong to raw foods that we primarily consider for diets (*i*) to (*iv*) (Supplementary Methods).

For the recommended daily levels of nutrient intake, we primarily referred to the Dietary Reference Intakes from the Institute of Medicine of the U.S. National Academies[28]. The data describe adequate calorie levels for daily activity and the lower and upper bounds of the recommended daily intake of 41 nutrients (Supplementary Table S1). Unlike our previous study[20], here we do not impose any lower bound of sodium intake, based on the assumption that the recommended sodium intake is readily achievable through the consumption of added salt, not necessarily through raw food consumption. In the case of diet (*iii*), vegan diet, we do not impose any lower bound of vitamin $B_{12}$ intake, because it is impossible to meet daily vitamin $B_{12}$ demands through the combinations of plant-derived foods only, without other nutritional conditions being compromised (Supplementary Table S1 and Supplementary Methods). For diets (*iv*) to (*vi*), we impose a very tight upper limit of methionine intake, as merely 10% more than the lower limit of the methionine intake (Supplementary Table S1).

**Nutritional fitness of foods across diets**

To specify food quality based on nutritional adequacy, we calculated the nutritional fitness of each food. Suppose that there is an ideal food that contains all necessary nutrients to meet, but not exceed, our daily nutrient demands. In this hypothetical scenario, consuming only this food will provide the optimal nutritional balance for our body. In the absence of such an ideal food, a realistic alternative would be to consume a set of foods, small in number, that still satisfies the nutritional recommendations. We identify these minimal food sets as the *irreducible food sets*, which were first introduced in our previous study[20]. For each diet, we generate irreducible food sets by solving optimisation problems with mixed-integer linear programming, as described in Supplementary Methods: each irreducible food set is a set of a small number of different foods that meet our daily nutrient demands in their entirety, and no irreducible food set is a superset of another set. The computed amounts of foods in irreducible food sets are provided in Supplementary Table S2. We only considered irreducible food sets that each has less than six different foods for diets (*i*) and (*iv*) and less than seven for the other diets. The total weight of the foods in each irreducible food set was limited to 4 kg, which reflects the practical limit of daily food consumption.

Foods with a frequent occurrence across irreducible food sets are likely to provide very balanced amounts of different nutrients, and would be classified as having high nutritional fitness (NF). In this regard, the NF of each food is provided by NF = log(*f*+1)/log(*N*+1), where *f* is the number of irreducible food sets that include the food and *N* is the total number of irreducible food sets [*N* = 52,957, 43,924, 20,713, 4,101, 1,053, and 5,225 for diets (*i*), (*ii*), (*iii*), (*iv*), (*v*), and (*vi*), respectively]. NF ranges from zero to one, and a large NF suggests that a food is nutritionally favourable. Note that *f* is capable of quantifying NF under the condition that the number of different foods comprising an irreducible food set is limited to a small number, as in this study. Otherwise, it may be difficult to estimate the true nutritional adequacy of foods solely from their *f* values. For example, a nutritionally poor food in an irreducible food set will be easily complemented by many other foods in the same food set, if the size of the food set is not sufficiently small.

All optimisation problems in this study were solved using Gurobi (v. 7.0.2) and IBM ILOG CPLEX (v. 12.4) solvers (Supplementary Methods).

**Key nutrients relevant to each food or diet**

To identify the individual nutrients responsible for high NFs of foods, we introduce a new quantity $\phi_{ij}$ for each pair of food *i* and nutrient *j*, as follows:

$$\phi_{ij} = \left\langle \frac{a_{ij} x_i}{\sum_k a_{kj} x_k} \right\rangle, \tag{1}$$

where $\alpha_{ij}$ is the density of nutrient $j$ in food $i$, $x_i$ is the weight of food $i$ to consume per day in a given irreducible food set, and $\langle \cdot \rangle$ is an average over all irreducible food sets that include the food $i$ (if $\sum_k \alpha_{kj} x_k = 0$ in any irreducible food set, this irreducible food set is excluded from the calculation). In other words, $\phi_{ij}$ represents the food $i$'s contribution to the total amount of the nutrient $j$ in an average irreducible food set. The value of $\phi_{ij}$ ranges from zero to one (Supplementary Data S1 and S2). If food $i$, on average, provides a certain proportion of nutrient $j$ in an irreducible food set, $\phi_{ij}$ takes the same value as that proportion. For example, if $\phi_{ij} = 0.7$, food $i$ supplies 70% of nutrient $j$, on average, in an irreducible food set. For any high-NF food $i$, we hence interpret nutrient $j$ with a large $\phi_{ij}$ as the main contributor to food $i$'s high NF. For a given value $\phi_{ij}$, we tested its statistical significance by calculating the one-sided $P$ value of how frequently $\phi_{i'j}$ of a randomly-chosen raw food $i'$ in the corresponding diet is greater than or equal to $\phi_{ij}$ (if the raw food $i'$ did not appear in any irreducible food sets, $\phi_{i'j}$ was treated as zero in this calculation).

For each diet, we also examined which nutrient is subjected to a risk of deficiency in its daily intake. Specifically, the contents of certain nutrients in irreducible food sets may merely meet and only marginally exceed the minimum daily requirement levels, even if food weights within the irreducible food sets are altered to maximise the intake of these nutrients. We consider these nutrients to be at risk of deficiency in that diet. For a given nutrient $j$, the following quantity $\theta_j$ is calculated:

$$\theta_j = \left\langle \frac{\max\left(\sum_k a_{kj} x_k\right) - L_j}{U_j - L_j} \right\rangle \text{ or } \theta_j = \left\langle \frac{\max\left(\sum_k a_{kj} x_k\right) - L_j}{\max\left(\sum_k a_{kj} x_k\right)} \right\rangle, \qquad (2)$$

where $L_j$ ($U_j$) is the lower (upper) bound of the recommended daily intake of nutrient $j$, max($\cdot$) is the maximum value among all multiple solutions with altered $x_k$'s in a given irreducible food set, and $\langle \cdot \rangle$ is an average over all irreducible food sets (Supplementary Methods). If nutrient $j$ has the upper bound of its recommended daily intake, we calculate the former $\theta_j$, and otherwise, the latter $\theta_j$. In other words, $\theta_j$ quantifies nutrient $j$'s maximally possible excess over its minimally required intake level in an irreducible food set, on average. The value of $\theta_j$ ranges from zero to one, and a small $\theta_j$ value indicates a risk of nutrient $j$'s deficiency.

**Dietary pattern design based on irreducible food sets**

To aid in the design of nutritionally recommendable dietary patterns, we prioritised irreducible food sets based on the overall NFs of foods within each irreducible food set. For the collection of irreducible food sets with the same, minimum size (e.g., irreducible food sets with four different foods in the case of a control diet), we sort these irreducible food sets in descending order by the largest NF of the foods in each irreducible food set. Next, for each collection of the irreducible food sets with the same largest NFs, we sort these irreducible food sets in descending order by the second largest NF of the foods in each irreducible food set. Sequentially, we repeat similar processes with the next largest NFs until every NF from each irreducible food set is examined for sorting. The top five irreducible food sets resulting from this sorting are designated as the *select dietary food sets*, which we perceive as guiding routes and benchmarks for the design of dietary patterns recommended for health. For each diet, we identified these select dietary food sets.

The select dietary food sets alone are based on the NFs of foods in irreducible food sets, and do not necessarily resemble actual dietary patterns in regular life. Therefore, among all the irreducible food sets, it would be valuable to search for food sets that are relatively close to regular dietary patterns. For example, one may consider the dietary patterns of people living in India, who are often familiar with vegetarian diets[29]. A recent study provides information on the major dietary patterns of an Indian population by principal component analysis of the consumed levels of individual food groups[30]. Each principal component corresponds to a dietary pattern. Based on this dataset, we selected irreducible food sets that closely match with each of the three dietary patterns (Supplementary Data S3): one is rich in cereals and savoury foods (cereal-savoury pattern), another is rich in fruits and vegetables (fruit-vegetable pattern), and the other is rich in animal foods (animal-food pattern). Specifically, in a given irreducible food set, the total amount of each food group was weighted by its factor loaded on a principal component, and these weighted quantities were summed across all food groups to give a component score of that irreducible food set. The top five irreducible food sets with the largest component scores were identified for each of the above three dietary patterns, across four different diets—control, lacto vegetarian, vegan, and methionine-restricted diets. Here, a lacto vegetarian diet, which reflects typical Indian vegetarianism[31, 32], excludes all egg products from the previously defined ovo-lacto vegetarian diet.

**Data availability**

All relevant data are available in the paper and Supplementary Information. Codes are available from the authors upon request.

**Results**

**Nutritional fitness of foods across diets**

We initially considered four diet styles with different plant-food consumption patterns: (*i*) control, (*ii*) ovo-lacto vegetarian, (*iii*) vegan, and (*iv*) methionine-restricted diets (see Methods). Here, diets (*i*) and (*iv*) are omnivorous, but the latter is far more plant-based. For diets (*i*) to (*iv*), we considered a physically active, 20-year-old male with regard to daily nutrient requirement levels (Supplementary Table S1). For each diet, we then obtained information on which foods can lead to good health outcomes by calculating the nutritional fitness (NF) of a food through the generation of nutritionally adequate food combinations (or irreducible food sets; see Methods). The value of NF for each food ranges from zero to one, and a large NF suggests that the food is nutritionally favourable (Figs 1a, b).

From our calculations, which foods then have the highest NFs in each diet? Noticeably, across all four diets, the most common foods with the highest NFs were almond, chia seeds, and cherimoya (Table 1 and Supplementary Data S1). Specifically, almond and chia seeds were always ranked in the top three foods regardless of diets [NF = 0.97 ± 0.01 and 0.93 ± 0.04 (avg. ± s.d.) for almond and chia seeds across all four diets, respectively]. Cherimoya was also ranked in the top three foods, except for the vegan diet where it still ranked in the top 5% of foods.

Before moving forward, we should stress that most foods can be grouped with each other according to their nutritional similarity[20], and the grouping results roughly correspond to the relative levels of food macronutrients, i.e., protein-rich, fat-rich, carbohydrate-rich, and low-macronutrient categories (Methods and Supplementary Data S1). According to this classification, the above high-NF foods, almond and chia seeds, belong to the fat-rich category, and cherimoya to the carbohydrate-rich category. One important issue here is whether the categories to which the foods belong play any role in NF-driven food prioritisation. Previously, we showed that using NFs to prioritise foods should only be conducted for foods that belong to the same category[20], because foods from different categories rather independently contribute to satisfying the overall nutritional requirements.

From this viewpoint, we sought high-NF foods within each separate category (Table 1 and Supplementary Data S1). We found that almond and chia seeds are the two highest-NF foods within the fat-rich category for every diet. In fact, all four diets commonly share two-thirds of the fat-rich category foods with NF > 0.5, and these shared, fat-rich and NF > 0.5 foods are all nuts and seeds (Supplementary Data S1). On the other hand, the protein-rich category exhibits a large variation in its high-NF foods across diets. Apart from the vegan diet with a minor role of the protein-rich category foods, the ovo-lacto vegetarian and omnivorous (control and methionine-restricted) diets are characterised by milk and fish products, respectively, as their highest-NF foods within the protein-rich category (Figs 1a, b and Table 1). At a finer level, a clear difference is even observed between the two omnivorous diets, control and methionine-restricted, in their highest-NF fish products: snapper and ocean perch are ranked in the top two fish products in the control diet, whereas roe and coho salmon are ranked in the top two in the methionine-restricted diet (Table 1 and Supplementary Data S1).

In the case of the carbohydrate-rich category, besides cherimoya with very high NF [NF = 0.87 ± 0.09 (avg. ± s.d.) across all diets], our focus was on frozen immature lima beans and frozen green peas. These two foods have high NFs in the vegan and ovo-lacto vegetarian diets, but not in the methionine-restricted diet. Frozen immature lima beans exhibit the highest and third-highest NFs within the carbohydrate-rich category in the vegan and ovo-lacto vegetarian diets, respectively, whereas the methionine-restricted diet shows the lowest NF for this food (NF = 0.28). Likewise, the methionine-restricted diet does not allow a high NF for frozen green peas (NF = 0.45), which are, in fact, consistently ranked in the top three NF foods within the carbohydrate-rich category across the other three diets. Overall, fruits tend to have higher NFs among the carbohydrate-rich foods, but this tendency is weaker in the vegan diet, which includes more grains, legumes, and others in the higher NF, carbohydrate-rich foods (Supplementary Data S1).

Lastly, in the low-macronutrient category, the ovo-lacto vegetarian and vegan diets have mushrooms as the highest-NF foods, such as maitake and ultraviolet-treated portabella mushrooms, whereas the control and methionine-restricted diets mainly have vegetables as higher-NF foods (Table 1 and Supplementary Data S1). In this study, we later revisit such diet- and category-specific results with a deeper nutritional analysis. A complete list of foods and their NFs in each diet are presented in Supplementary Data S1.

Another interesting work is to search for foods low in calories but still with high NFs. These foods might be comparable to, yet distinct from, foods with high micronutrient levels per calorie, which have been suggested to benefit weight and glycaemic control and cardiovascular risk management[33, 34]. We first observed that the calorie level of each food clearly follows a bimodal distribution (Supplementary Fig. S1). In other words, every food in our study can be classified as having either a high or low calorie level by a natural threshold from this bimodal distribution (Supplementary Fig. S1). Supplementary Data S1 provides a list of low-calorie foods with high NF in each diet (NF > 0.7), such as cherimoya and dandelion greens in every diet.

**Key nutrients relevant to each food or diet**

The NFs of foods in our study were found to have diet-dependent characteristics. The next step is to deeply examine the individual nutrients responsible for the NFs of foods in the context of different diets. For example, in the case of vegetarian diets, what particular nutrients are responsible for mushrooms having the highest NFs among the low-macronutrient category foods? To identify these key nutrients, we defined the quantity $\phi_{ij}$ that measures nutrient $j$'s dependency on food $i$ in irreducible food sets (0 ≤ $\phi_{ij}$ ≤ 1; see Methods). A large $\phi_{ij}$ suggests that nutrient $j$ is the main contributor to the NF of food $i$.

The analysis of $\phi_{ij}$ shows that vitamin D is the main contributor to the mushrooms' high NF in vegetarian diets (Fig. 1c and Supplementary Data S1). For both the ovo-lacto vegetarian and vegan diets, every type of mushrooms with NF > 0.75 serves as a good source of vitamin D with $\phi_{ij}$ > 0.9 ($P$ ≤ 0.01; Methods and Supplementary Data S1). It is worthwhile to note that, across all four diets, the low-macronutrient category foods other than mushrooms are mainly contributing vitamin K and vitamin A, not vitamin D (Supplementary Data S1).

In the control and methionine-restricted diets, most of the vitamin D can be provided by fish products, which substantially contribute to the intake of vitamin $B_{12}$, selenium, essential amino acids, and choline as well (Supplementary Data S1). Vitamin D is one of the key players for the NFs of finfish being generally higher than the NFs of the other protein-rich foods (e.g., milk, pork, beef, and poultry) in these omnivorous diets, as demonstrated through the vitamin D's $\phi_{ij}$ values across the protein-rich foods (Supplementary Data S1).

Regarding a vegan diet, an important issue is to explore the food sources of essential amino acids, as this diet lacks protein-rich category foods in most irreducible food sets. Remarkably, we found that the aforementioned, frozen immature lima beans serve as an excellent amino acid source in the vegan diet (Fig. 1d), providing lysine ($\phi_{ij}$ = 0.76), isoleucine ($\phi_{ij}$ = 0.76), valine ($\phi_{ij}$ = 0.72), methionine ($\phi_{ij}$ = 0.62), and so forth ($P$ = 0.002; Methods and Supplementary Data S1). Frozen green peas play a similar role in the vegan diet; they provide a large portion of lysine ($\phi_{ij}$ = 0.57), methionine ($\phi_{ij}$ = 0.54), and threonine ($\phi_{ij}$ = 0.50) in the vegan diet ($P$ ≤ 0.007; Methods and Supplementary Data S1). These high methionine contributions may partially explain our prior observation that both frozen immature lima beans and green peas have lower NFs in the methionine-restricted diet.

Lastly, foods with consistently high NF across all four diets, almond, chia seeds, and cherimoya, were found to have very large $\phi_{ij}$ values ($\phi_{ij}$ > 0.9 and $P$ ≤ 0.005) for two essential fatty acids, linoleic acid (almond) and α-linolenic acid (chia seeds and cherimoya) (Figs 1e, f and Supplementary Data

S1). Together, Figures 1c-f present the examples of nutrients with large $\phi_{ij}$ values from several high-NF foods.

In Methods, we have already pointed out that a vegan diet hardly meets the daily vitamin $B_{12}$ demands. Likewise, each of the four diets may involve a set of nutrients at risk of deficiency in their daily intake levels. Hence, it will be important to recommend consumption of foods that can mitigate against potential nutrient deficiencies in each diet. To this end, we evaluated the susceptibility of nutrients to deficiency in each diet by leveraging information from irreducible food sets (Methods), and then found foods with high $\phi_{ij}$ for the potentially deficient nutrients.

Surprisingly, we found that all four diets commonly involve a risk of choline, vitamin $B_6$, and vitamin E deficiency (Table 2). In addition, vegetarian and methionine-restricted diets involve selenium as a micronutrient of potential deficiency. Selenium is a constituent of selenoproteins[35], and we indeed observed a very positive correlation between selenium and protein levels across foods[20] ($r$ = 0.74 and $P$ = 1.0 ×10$^{-41}$; Supplementary Methods). Therefore, vegetarian and methionine-restricted diets may be of concern based on the possible shortage of selenium intake. A vegan diet among them, as well as a methionine-restricted diet, is of further concern due to the risk of methionine deficiency (Table 2).

Recommended foods with high $\phi_{ij}$ for the above nutrients of concern include fish products for choline, selenium, and methionine, and in a vegan diet, frozen immature lima beans for choline, selenium, methionine, and vitamin $B_6$ (Supplementary Data S1). We also recommend almond and cherimoya for vitamin E ($\phi_{ij}$ > 0.7 and $P$ ≤ 0.002) and vitamin $B_6$ ($\phi_{ij}$ > 0.5 and $P$ ≤ 0.03), respectively (Supplementary Data S1). We suggest that deeper analyses into these nutrients at risk of deficiency may be warranted when the prioritisation of diet-specific foods is of interest.

**Dietary pattern design based on irreducible food sets**

NF offers a way to recommend individual foods for nutritionally adequate dietary plans. On the other hand, one can recommend nutritionally adequate food combinations *per se* (i.e., irreducible food sets), which we call the select dietary food sets (see Methods). They are a collection of five representative irreducible food sets in each diet, which tend to include many high-NF foods inside.

Supplementary Data S1 presents information on the select dietary food sets and food weights to consume. In the case of the vegan diet, the select dietary food sets commonly include chia seeds, almond, frozen immature lima beans, and ultraviolet-treated portabella mushroom. In addition, the following foods are included in the select dietary food sets, and each food serves as an alternative to another: dandelion greens, red leaf lettuce, sun-dried hot chilli peppers, paprika, and young green onions (tops only) (Supplementary Data S1). For the vegan diet, we hence recommend all those foods together, along with vitamin $B_{12}$ supplements, as no irreducible food set in the vegan diet satisfies daily vitamin $B_{12}$ requirements.

In the case of the ovo-lacto vegetarian diet, the select dietary food sets are composed of almond, cherimoya, maitake mushroom, milk, and frozen green peas. Among them, cherimoya, maitake mushroom, and frozen green peas can be replaced simultaneously by plums, shiitake mushroom, and chia seeds. In the case of the methionine-restricted diet, the select dietary food sets are composed of almond, chia seeds, cherimoya, roe, and red cabbage, while cherimoya can be replaced by plantains and corn (or potato) flour. Lastly, the control diet involves the select food sets with almond, cherimoya, snapper (or alternatively, ocean perch), and dandelion greens. In this case, dandelion greens can be replaced by red cabbage, Swiss chard, and broccoli raab (Supplementary Data S1).

Together, the select dietary food sets show promise for formulating nutritionally adequate dietary patterns based on the respective foods' NFs. Apart from the select dietary food sets, one may seek the food sets relatively close to actual dietary patterns in regular life, such as those of people living in India, who are often familiar with vegetarian diets[29]. For a given diet, we selected a collection of five irreducible food sets with the closest match to each of the following three Indian dietary patterns (Methods): one is rich in cereals and savoury foods (cereal-savoury pattern), another is rich in fruits and vegetables (fruit-vegetable pattern), and the other is rich in animal foods (animal-food pattern).

The resulting food sets are found in Supplementary Data S3. They involve some food items other than in the select dietary food sets, such as rice in the cereal-savoury pattern of the control and vegan diets, grapefruit and melons in the fruit-vegetable pattern of the lacto vegetarian and vegan diets, and egg whites in the animal-food pattern of the control diet (we consider a lacto vegetarian diet instead of an ovo-lacto vegetarian diet, because typical Indian vegetarianism is lacto vegetarianism[31, 32]). As expected, the lacto vegetarian and vegan diets do not have any irreducible food sets of the animal-food pattern (Supplementary Data S3). Considerations of more regional factors such as local food repertoires would be warranted for further improvements in our analysis.

**Application to personalised food recommendations**

Thus far, this work has provided an overview of nutritionally recommendable foods in semi- to strict vegetarian diets compared to a control diet. Now, we extend our analysis to two hypothetical scenarios of highly personalised, plant-based methionine-restricted diets with limited animal-food consumption. The personalised case I pertains to a 61-year-old male with low physical activity, and the personalised case II pertains to a physically active 58-year-old female. These two people also differ in their dietary preferences (see Methods for more details).

Through the calculation of NF as shown in Figs 2a, b, we found that high-NF foods in both cases I and II are generally reminiscent of our former results for the ovo-lacto vegetarian and vegan diets, with an exception of the protein-rich category's high-NF foods that are clearly distinct from the former ones (Table 3 and Supplementary Data S2).

Specifically, in a similar fashion to the other diets, almond and chia seeds are the two highest-NF foods with NF > 0.85 within the fat-rich category for both cases I and II. In the carbohydrate-rich category, cases I and II have frozen immature lima beans and cherimoya, respectively, as the highest-NF foods with NF > 0.7. In the low-macronutrient category, ultraviolet-treated portabella and maitake mushrooms, squash, and dandelion greens are ranked in the top four foods for both cases I and II.

In contrast to the above results, the protein-rich category of both cases I and II includes dried whitefish and dried smelt as the top-NF foods with NF > 0.8, which are not foods allowed by the previous vegetarian diets and have low NF < 0.5 in the previous control and methionine-restricted diets. Dried whitefish and smelt exhibit very high $\phi_{ij}$ > 0.97 with vitamin B$_{12}$ in cases I and II ($P \leq 0.006$;

Fig. 2c and Supplementary Data S2), suggesting that both foods are an excellent source of vitamin $B_{12}$ in these diets.

Despite the overall similarity between cases I and II, we observed some differences in their recommended foods. For example, in case I, herring shows a moderate level of NF with NF = 0.58, but does not even reach such NF levels in case II (Figs 2a, b), because any irreducible food sets that we examined in case II do not contain herring. It is noteworthy that the irreducible food sets do not include dairy products as well, in both cases I and II, although these two cases allow milk and cheese consumption. In these two cases, among animal-derived foods, fish would be a better dietary choice than dairy products, to meet nutritional requirements.

Consistent with the aforementioned results of the vegetarian and methionine-restricted diets, our analysis reveals that the dietary patterns in cases I and II involve a risk of choline, vitamin $B_6$, vitamin E, and selenium deficiency (Table 4 and Methods). Recommended foods with high $\phi_{ij}$ for those nutrients of concern include frozen immature lima beans for choline ($\phi_{ij} > 0.6$ and $P \leq 0.006$) and vitamin $B_6$ ($\phi_{ij} > 0.45$ and $P \leq 0.04$), almond for vitamin E ($\phi_{ij} > 0.77$ and $P \leq 0.002$), cherimoya for vitamin $B_6$ ($\phi_{ij} > 0.45$ and $P \leq 0.04$), and dried smelt for selenium ($\phi_{ij} > 0.35$ and $P \leq 0.04$) (Figs 2d, e and Supplementary Data S2).

To recommend food sets that help design healthy dietary patterns for cases I and II, we obtained the select dietary food sets, as defined above (Methods). Supplementary Data S2 presents information on the select dietary food sets and food weights to consume in cases I and II. In case I, the select dietary food sets commonly include dried whitefish, hubbard squash, almond, chia seeds, and ultraviolet-treated portabella mushroom. In addition, the following foods are included in the select dietary food sets, and each food serves as an alternative to another: dandelion greens, broccoli raab, watercress, red cabbage, and Swiss chard (Supplementary Data S2). In case II, one of the select dietary food sets is composed of dried smelt, zucchini, almond, ultraviolet-treated portabella mushroom, cherimoya, and sweet potato. Among them, cherimoya and sweet potato can be replaced simultaneously by dandelion greens and corn flour (or alternatively, garlic powder) (Supplementary Data S2).

In summary, our approach can be used for a systematic investigation of highly personalised, plant-based diets. These personalised diet results are largely reflective of our previous vegetarian and methionine-restricted diet results. Nevertheless, we observe clearly distinct, individualised features, particularly from the protein-rich food items for vitamin $B_{12}$ consumption.

**Discussion**

In this study, we have explored semi- to strict vegetarian, and control omnivorous diets through the systematic analysis of large-scale food and nutritional data. Our study offers a global and unbiased view of the nutritional structure of vegetarian diets, as well as enables the discovery of unconventional knowledge of foods and nutrients. Nutritional fitness, which gauges the quality of a raw food according to its nutritional density and balance, appears to be useful to prioritise recommended foods for vegetarian health, notwithstanding vegetarians' limited food repertoire compared to a standard omnivorous diet. Extending our analysis to two highly personalised, semi-vegetarian diets has revealed the applicability of nutritional fitness towards personalised food recommendations[36]. People of different diets, ages, genders, body compositions, health states, and physical activity levels can obtain their condition-specific food information through our method, by simply adjusting the food repertoire and required calorie and nutrient intakes when generating irreducible food sets (Supplementary Methods). The resulting irreducible food sets allow one to compute the nutritional fitness and identify the key relevant nutrients, as well as the nutrients susceptible to deficiency. However, this study is limited to the analysis of mainly raw foods. Extending the analysis beyond raw foods to cooked foods is necessary to explore the comprehensive repertoire of the foods we consume daily, as well as their temperature-sensitive nutrients that may be destroyed by cooking, such as vitamins A, C, and E, thiamin, and folate[37].

As a step forward, more realistic applications will be achievable if the concepts presented here are judiciously incorporated with other practical factors. Considerations of food taste and cultural, regional, seasonal, and financial factors in our analysis may improve the practical applicability of our methods in daily healthcare experiences[38, 39]. In particular, food taste or gastronomy may become a major factor in future studies[18, 40]; currently, our select dietary food sets for dietary pattern designs tend to suggest consuming very large amounts of certain foods, like squash as a source of various vitamins and α-linolenic acid in the aforementioned personalised dietary cases (Fig. 2f and Supplementary Data S2). Such a tendency is generally observed from the weights of foods in irreducible food sets, as indicated in Supplementary Table S2. Therefore, these strong preferences for particular foods in the select dietary food sets reflect the nature of irreducible food sets that have been optimised only on a nutritional basis, without considering the gastronomic aspects. Nevertheless, these results offer a primary insight into nutritionally beneficial foods, but warrant further realistic sophistication by gastronomic consideration. An interesting avenue is to incorporate regional dietary patterns into our analysis, as demonstrated above with the preliminary results for Indian dietary patterns.

Finally, we envision that nutritional studies adopting our systematic approach will be at the forefront of designing condition-specific, dietary interventions to promote wellness[24, 25], and may provide a useful basis for nutritional policy making, nutrition education, and food marketing[41, 42]. Moreover, the combination of information technology, healthcare services, and our approach will enforce the accessibility of nutritionally recommendable food items to the general public[43, 44].

## Acknowledgements

This work was supported by the National Research Foundation of Korea (NRF) Grant NRF-2015R1C1A1A02037045 funded by the Korean Government (MSIP) (S.K. and P.-J.K.). This work was also supported by the ICTP through the OEA-AC-98 (S.K.).

**Figures**

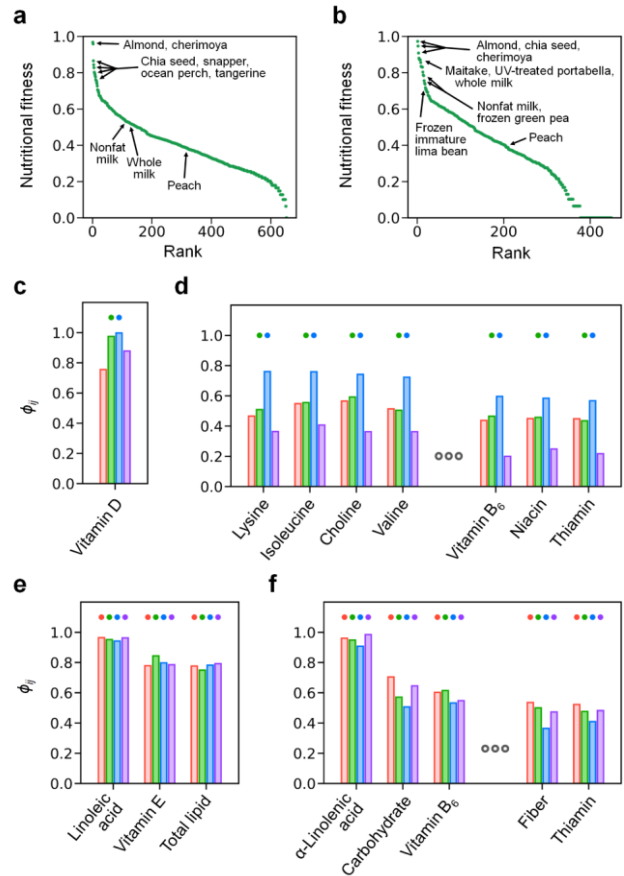

**Figure 1. Nutritional fitness (NF) and underlying nutrients across diets.** We here consider four different diets, control, ovo-lacto vegetarian, vegan, and methionine-restricted diets. For the definition of these diets, refer to the main text. (**a**) NFs of foods in a control diet, sorted in descending order. (**b**) NFs of foods in an ovo-lacto vegetarian diet, sorted in descending order. Given food *i* and nutrient *j*, $\phi_{ij}$ indicates the contribution of nutrient *j* to the NF of food *i*, as described in the main text. The specific value of $\phi_{ij}$ depends on diets. From each of the following foods in (**c**) to (**f**), $\phi_{ij}$ is plotted for the nutrients with $\phi_{ij} > 0.5$ in any of the above considered diets: (**c**) maitake mushroom, (**d**) frozen immature lima beans, (**e**) almond, and (**f**) cherimoya. For each nutrient in (**c**) to (**f**), four $\phi_{ij}$ values from the four considered diets are presented sequentially in the following order: control (red), ovo-lacto vegetarian (green), vegan (blue), and methionine-restricted (purple) diets. In a given diet, if the NF of the food is > 0.7, the top of $\phi_{ij}$ from that diet is dotted. In (**d**) and (**f**), some nutrients with intermediate $\phi_{ij}$ values are omitted for visual clarity, and all of the data can be found in Supplementary Data S1.

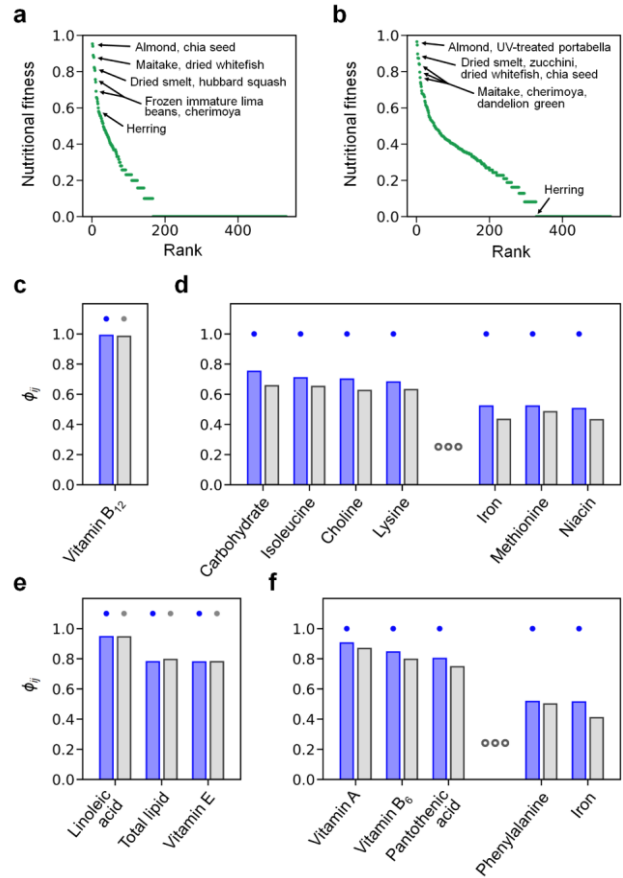

**Figure 2. Highly personalised, mainly plant-based diets.** We here consider the following two personalised cases: 61-year-old male (case I) and 58-year-old female (case II), who practise a plant-based, methionine-restricted diet, and consume a limited amount of animal products. Cases I and II are different in their physical conditions and specific food intakes, as described in the main text. (**a**) NFs of foods in case I, sorted in descending order. (**b**) NFs of foods in case II, sorted in descending order. Given food *i* and nutrient *j*, $\phi_{ij}$ indicates the contribution of nutrient *j* to the NF of food *i*, as described in the main text. The specific value of $\phi_{ij}$ depends on cases I and II. From each of the following foods in (**c**) to (**f**), $\phi_{ij}$ is plotted for the nutrients with $\phi_{ij} > 0.5$ in either case I or II: (**c**) dried smelt, (**d**) frozen immature lima beans, (**e**) almond, and (**f**) hubbard squash. For each nutrient in (**c**) to (**f**), two $\phi_{ij}$ values from cases I and II are presented sequentially (case I in blue and case II in grey). In case I or II, if the NF of the food is > 0.7, the top of $\phi_{ij}$ from that case is dotted. In (**d**) and (**f**), some nutrients with intermediate $\phi_{ij}$ values are omitted for visual clarity, and all of the data can be found in Supplementary Data S2.

## Tables

**Table 1. Foods with high nutritional fitness (NF) across diets.** We here consider four different diets, control, ovo-lacto vegetarian, vegan, and methionine-restricted diets. For the definition of these diets, refer to the main text. In each food category, we sort foods by their highest NFs among the four diets and list only the top 5 cases (in the protein-rich category, several milk products with similar highest NFs are listed together in the same row for visual clarity). Each of these foods has NF > 0.7 in at least one diet. For each food, we specify the diets that give NF > 0.7 (C, control; O, ovo-lacto vegetarian; V, vegan; M, methionine-restricted; the specific value of NF is presented in parentheses beside each diet).

| Food category | Food | Diet with high nutritional fitness |
|---|---|---|
| Protein-rich | Whole milk | O (0.86) |
| | Nonfat dry milk, reduced fat milk, 1%-fat milk | O (0.83) |
| | Snapper | C (0.83) |
| | Ocean perch | C (0.80) |
| | Roe | C (0.70), M (0.79) |
| Fat-rich | Almond | C (0.97), O (0.97), V (0.97), M (0.99) |
| | Chia seed | C (0.87), O (0.95), V (0.98), M (0.93) |
| | Dried pumpkin and squash seed kernels | C (0.84), O (0.87), V (0.87) |
| | Pork separable fat | C (0.80) |
| | Dried black walnut | O (0.71), V (0.77) |
| Carbohydrate-rich | Cherimoya | C (0.96), O (0.91), V (0.72), M (0.89) |
| | Frozen immature lima bean | O (0.72), V (0.85) |
| | Frozen green pea | O (0.76), V (0.80) |
| | Tangerine | C (0.76), V (0.78), M (0.77) |
| | Full-fat soy flour | V (0.76) |
| Low-macronutrient | Ultraviolet-treated portabella | O (0.87), V (0.94) |
| | Maitake | O (0.88), V (0.87) |
| | Dried shiitake | O (0.87) |
| | Red cabbage | C (0.74), O (0.71), M (0.83) |
| | Chanterelle | O (0.80), V (0.76) |

**Table 2. Nutrients at risk of deficiency across diets.** We here consider four different diets, control, ovo-lacto vegetarian, vegan, and methionine-restricted diets. For the definition of these diets, refer to the main text. In a given diet, each nutrient is assigned $\theta$, and low $\theta$ indicates a risk of the nutrient deficiency in that diet (see Methods). We list the nutrients that have $\theta \leq 0.15$ in at least one diet. For each nutrient, we specify the diets that give $\theta \leq 0.15$ (C, control; O, ovo-lacto vegetarian; V, vegan; M, methionine-restricted; the specific value of $\theta$ is presented in parentheses beside each diet). Here, we omit rather trivial information, i.e., the risk of vitamin $B_{12}$ deficiency in a vegan diet (see the main text) and the risk of methionine deficiency in a methionine-restricted diet.

| Nutrient | Diet with low $\theta$ |
|---|---|
| Choline | C (0.03), O (0.03), V (0.04), M (0.03) |
| Vitamin $B_6$ | C (0.05), O (0.04), V (0.03), M (0.05) |
| Vitamin E | C (0.04), O (0.04), V (0.04), M (0.05) |
| Selenium | O (0.13), V (0.12), M (0.09) |
| Methionine | V (0.12) |
| Zinc | M (0.14) |
| Protein | V (0.14) |

**Table 3. Foods with high nutritional fitness (NF) in highly personalised, mainly plant-based diets.** We here consider the following two personalised cases: 61-year-old male (case I) and 58-year-old female (case II), who practise a plant-based, methionine-restricted diet, and consume a limited amount of animal products. Cases I and II are different in their physical conditions and specific food intakes, as described in the main text. In each food category, we show only the foods that have NF > 0.7 in either case I or II. For each food, we specify the case I or II that gives NF > 0.7 (I, case I; II, case II; the specific value of NF is presented in parentheses beside each case).

| Food category | Food | Diet with high nutritional fitness |
|---|---|---|
| Protein-rich | Dried smelt | I (0.82), II (0.90) |
| | Dried whitefish | I (0.88), II (0.87) |
| | Dried chum salmon | II (0.74) |
| | Common octopus | II (0.72) |
| Fat-rich | Almond | I (0.95), II (0.96) |
| | Chia seed | I (0.94), II (0.87) |
| | Dried pumpkin and squash seed kernels | I (0.82), II (0.84) |
| Carbohydrate-rich | Cherimoya | II (0.80) |
| | Frozen immature lima bean | I (0.76) |
| Low-macronutrient | Ultraviolet-treated portabella | I (0.89), II (0.95) |
| | Maitake | I (0.88), II (0.84) |
| | Zucchini | II (0.87) |
| | Hubbard squash | I (0.81) |
| | Dandelion green | I (0.74), II (0.77) |
| | Chanterelle | II (0.73) |

**Table 4. Nutrients at risk of deficiency in highly personalised, mainly plant-based diets.** We here consider the following two personalised cases: 61-year-old male (case I) and 58-year-old female (case II), who practise a plant-based, methionine-restricted diet, and consume a limited amount of animal products. Cases I and II are different in their physical conditions and specific food intakes, as described in the main text. In case I or II, each nutrient is assigned $\theta$, and low $\theta$ indicates a risk of the nutrient deficiency in that case (see Methods). We list the nutrients that have $\theta \leq 0.15$ in either case I or II. For each nutrient, we specify the case I or II that gives $\theta \leq 0.15$ (I, case I; II, case II; the specific value of $\theta$ is presented in parentheses beside each case). Here, we omit rather trivial information of a risk of methionine deficiency in these methionine-restricted diets.

| Nutrient | Diet with low $\theta$ |
| --- | --- |
| Choline | I (0.02), II (0.02) |
| Vitamin E | I (0.03), II (0.03) |
| Vitamin $B_6$ | I (0.04), II (0.04) |
| Selenium | I (0.12), II (0.15) |